# Chromatin Laser Imaging Reveals Abnormal Nuclear Changes for Early Cancer Detection

Yu-Cheng Chen[1,2], Qiushu Chen[1], Xiaotain Tan[1], Grace Chen[3], Ingrid Bergin[4], Muhammad Nadeem Aslam[5*], and Xudong Fan[1*]

[1]Department of Biomedical Engineering, University of Michigan,

1101 Beal Ave., Ann Arbor, MI, 48109, USA

[2]School of Electrical and Electronics Engineering, Nanyang Technological University,

50 Nanyang Ave, 639798, Singapore

[3]Comprehensive Cancer Center, University of Michigan, Ann Arbor, MI 48109, USA

[4]Unit for Laboratory Animal Medicine, University of Michigan, Ann Arbor, MI 48109, USA

[5]Department of Pathology, University of Michigan,

1301 Catherine Street, Ann Arbor, MI 48109, USA

*Correspondence: xsfan@umich.edu, mnaslam@med.umich.edu

**Abstract**

We developed and applied rapid scanning laser-emission microscopy (LEM) to detect abnormal changes in cell nuclei for early diagnosis of cancer and cancer precursors. Regulation of chromatins is essential for genetic development and normal cell functions, while abnormal nuclear changes may lead to many diseases, in particular, cancer. The capability to detect abnormal changes in "apparently normal" tissues at a stage earlier than tumor development is critical for cancer prevention. Here we report using LEM to analyze colonic tissues from mice at-risk for colon cancer (induced by a high-fat diet) by detecting pre-polyp nuclear abnormality. By imaging the lasing emissions from chromatins, we discovered that, despite the absence of observable lesions, polyps, or tumors under stereoscope, high-fat mice exhibited significantly lower lasing thresholds than low-fat mice. The low lasing threshold is, in fact, very similar to that of adenomas and is caused by abnormal cell proliferation and chromatin deregulation that can potentially lead to cancer. Our findings suggest that conventional methods, such as colonoscopy, may be insufficient to reveal hidden or early tumors under development. We envision that this work will provide new insights into LEM for early tumor detection in clinical diagnosis, and fundamental biological and biomedical research of chromatin changes at the biomolecular level of cancer development.

Laser-emission microscopy (LEM) is an emerging imaging technology for biomedical research and medical diagnosis[1-3]. In LEM, a piece of tissue (either frozen or formalin-fixed paraffin-embedded (FFPE)) stained with dyes is sandwiched between two mirrors that form a laser cavity. External excitation is scanned over the tissue and the laser emission from the staining dyes is detected and used to analyze tissues. Fundamentally different from fluorescence, laser emission has threshold behavior, narrow spectral linewidth, and strong intensity[4-13], leading to ultrasensitive detection, superior image contrast[2,14], and high spectral/spatial resolution[14,15]. To date, LEM has been applied to various types of human normal and cancerous tissues, including lung, breast, colon, and stomach[2,3]. The corresponding imaging protocol has also been developed and shown to be highly compatible with the sample preparation and staining routines in pathological laboratories[3]. Through extensive research, we have found that cancer tissues (or cells) have a much lower lasing threshold than normal tissues (or cells) when their nuclei are stained with dyes (such as YOPRO), which is attributed to the higher gain in the nucleus-staining dyes in the cancer cells that are more active and undergo higher cell proliferation and chromatin condensation[2,3]. This phenomenon can be exploited to differentiate cancer and normal tissues with a high sensitivity and specificity. Indeed, in our recent work using Stage I/II human lung cancer tissues, an area under the Receiver Operating Characteristics (ROC) curve of 0.998 was achieved[2]. However, in reality cancer or cancer precursors may already be in progress at the biomolecular level (e.g., at the DNA level), much earlier than the structural and morphological changes (such as appearance of colonic polyps or tumors) that are usually detected by traditional microscopy. Therefore, given the unique LEM's capability to detect cellular (chromatin) deregulations, we hypothesize that LEM may be able to pick up the signatures of cancer or cancer precursors at an earlier stage, which is critical for cancer treatment and prevention.

Among all cancer types, colorectal cancer is one of the most common cancers[16] and top third leading cause of cancer-related deaths worldwide with estimated 51,000 deaths in 2018 in the United States. Although overall incidence of colorectal cancer is decreasing in older adults, the incidence has been increasing in the United States among adults younger than 55 years old with the increase confined to white men and women and most rapid for metastatic disease[17]. For people at risk for colorectal cancer, colonoscopy screening is usually performed to examine the presence of polyps (precursors for colon cancer) or cancerous lesion. These screening colonoscopies may turn out to be negative when either there are flat or depressed lesions or very small lesions that may not be detected without using any dye during the scoping. Some other precancerous lesions that are difficult to resect, such as sessile serrated polyps, also add to this list and cause increased colonic tumorigenesis. Improving prevention surveillance strategies that can improve detection will be a hallmark of lowering precancerous lesions. The adenoma detection rate is one of the most important features that can be improved to decrease the future development of colon cancer[18]. However, it is known that precancerous changes may be in progress even when a patient is only of 20-30 years of age, well before the appearance of polyps[19,20]. Thus the capability to detect abnormal changes in *normal* colon tissues at a stage earlier than polyps is extremely important for colon cancer prevention and treatment. Precancerous polyps tend to grow slow, but fastest growing polyps or cancers may double in 20 to 122 weeks[21,22]. Therefore, there is an unmet need to introduce newer technologies which can complement current strategies being used to improve the detection rates.

As the first step towards early diagnosis of colon cancer or cancer precursor, in this work we used colonic samples from C57BL/6 mice at risk for colon cancer to test the LEM's capability in detecting pre-polyp nuclear abnormality. These C57BL/6 mice were used and well characterized in a previously completed study to understand the risk for colon cancer due to lifelong intake of the high-fat Western style diet (HFWD)[23]. AIN-76A was used as low-fat control diet. In that study, a total of 40 C57BL/6 mice (20 males and 20 females) were maintained for 15 months on either the high-fat or low-fat diets (high-fat: 10 males and 10 females; low-fat: 10 males and 10 females. See Fig. 1a). They were euthanized at the end of the 15-month period and the histological sections of colons were evaluated and characterized microscopically. Published data has shown that the mice (and particularly the female mice) fed with a high-fat diet have a higher incidence of colon polyps (adenomas) than those fed with a low-fat diet (20% incidence in the high-fat mice - all female vs. 0% incidence in the low-fat mice)[23]. A follow-up time course study with a study duration of 18 months demonstrated that even the low-fat mice developed polyps, 36% as compared to 47% in the high-fat group after 18 months of dietary intervention[24]. Additionally, 16% of these precursor lesions turned out to be invasive adenocarcinomas when characterized histologically. The majority of these mice who developed polyps were also females, regardless of their diet[24]. Therefore, those 40 mice from the 15-month study can serve as an excellent at-risk model to test the LEM capability in detecting nuclear changes prior to polyp formation.

Here we employed LEM to re-analyze the FFPE colonic tissues from the aforementioned high-fat and low-fat 15-month mouse model. We found, surprisingly, that *despite the absence of observable lesions, polyps, or tumors under stereoscope*, the high-fat fed mice, and particularly, the high-fat fed female mice, exhibited significantly low lasing thresholds compared to the low-fat fed counterparts, indicative of the chromatin deregulation in those "apparently normal" high-fat fed mouse colon tissues. This finding agrees well with the results from our previous published studies[23,24] as well as the generally accepted opinion regarding diet-induced or obesity-related colon cancer incidence[25-27]. Therefore, our study may help detect early changes in nuclei and assist in better understanding at the biomolecular level regarding how colon cancer evolves. In addition, by working together with the conventional pathological methods and tools (e.g., H&E and IHC), we envision that LEM may provide a new paradigm for earlier diagnosis of cancers and cancer precursors.

## Results

In this study, 50 normal colon tissues (Fig. 1a) labeled with nucleic acid probe (YOPRO) were sandwiched in an FP microcavity while a high-repetition rate excitation laser was scanned to build a rapid-scanning LEM for large area diagnosis, as illustrated in Fig. 1b. Here "normal tissues" are referred to as those with no detectable polyps or tumors by using stereomicroscope. Based on conventional microscopy (H&E), the morphology of cell nuclei remain similar among all low-fat and high-fat normal colon tissues (left panel in Fig. 1b). However, laser emissions from the "apparently normal" cells/tissues may turn out to be significantly different between high-fat and low-fat dietary treatment, owing to chromatin deregulations and altered DNA concentrations (right panel in Fig. 1b).

As the baseline, we first investigated the lasing thresholds of normal colon tissues extracted from 10 individual new-born mice (1-month mice, 5 males and 5 females) without any dietary treatment (i.e., low-fat diet). The lasing spectra under various pump energy densities are given in Fig. 2a, showing highly stable lasing peak positions due to the spacer used in the cavity that fixes the cavity length[3]. The spectrally integrated laser output as a function of pump energy density extracted from the laser spectra is plotted in Fig. 2b and the lasing threshold of 11 μJ/mm$^2$ is obtained. Fig. 2c further compares the lasing threshold for the individual ten mice. For each mouse, 10 cells were selected randomly from the epithelial part of the colon tissue along the colon crypt.

Obviously, the lasing thresholds for all the samples are relatively low, mostly below 40 μJ/mm$^2$, due to the high cell proliferation and thus chromatin condensation in new born mice[28]. In addition, there is no significant difference in the lasing threshold between males and females, which is further verified by the histogram in Fig. 2d.

Next, we studied the lasing thresholds of normal colon tissues from 40 15-month mice (adult mice) with different dietary treatments. These 40 mice were classified into 4 subgroups, low-fat male, low-fat female, high-fat male, and high-fat female mice, each of which subgroups had 10 mice (see also Fig. 1a). The corresponding H&E images are provided in Supplementary Fig. 2. Fig. 3a shows an exemplary lasing spectrum from a low-fat male colon tissue under a pump energy density of 50 μJ/mm$^2$. For comparison, Fig. 3b shows an exemplary lasing spectrum extracted from a high-fat female colon tissue under a pump energy density of 50 μJ/mm$^2$. The corresponding spectrally integrated laser output as a function of pump energy density extracted from the lasing spectra in Figs. 3a and 3b is plotted in Fig. 3c. The solid lines are the linear fit above the lasing threshold, which is 40 and 14 μJ/mm$^2$, respectively. When pumped at 30 μJ/mm$^2$, we can clearly see that only the high-fat female tissue is able to lase due to higher nucleic acids concentrations (i.e., more gain from YOPRO) in the cell nucleus, as presented in Fig. 3d. Comparison of the confocal fluorescence microscopic images in Fig. 3e shows no significant difference between the low-fat female and high-fat male colon tissues. In contrast, a huge difference can be observed when viewed under laser-emission images. In Fig. 3f we further show representative lasing emission images taken from the same low-fat male and high-fat female colon tissues when pumped at 30 μJ/mm$^2$. No lasing emission is seen from the low-fat male colon tissues, whereas strong lasing emission with a superior contrast against the surrounding background is seen from the cells in high-fat female colon tissue. When the pump energy density is increased to 70 μJ/mm$^2$, multiple lasing cells can be seen for both low-fat male and high-fat female colon tissue in Fig. 3g, but the high-fat female tissue has more lasing cells and stronger lasing emission from each of those lasing cells.

In order to understand whether high-fat diet plays a key role in the laser emissions of both genders, in Fig. 4 we performed the lasing thresholds of colon tissues collected from the 20 15-month male mice. Fig. 4a shows the lasing thresholds of 10 individual low-fat male mice where the lasing thresholds seem to have a wide range of distribution from 40 μJ/mm$^2$ to 130 μJ/mm$^2$. In comparison, Fig. 4b shows the lasing thresholds of 10 high-fat male mice whose lasing threshold range is lowered to 30 µJ/mm$^2$ to 70 µJ/mm$^2$. The histogram of all lasing thresholds collected from both low-fat and high-fat is provided in Fig. 4c and the corresponding Receiver Operating Characteristics (ROC) curve is given in the inset. The area under the ROC curve is only 0.71, which indicates a low sensitivity and specificity in terms of the dietary effect on males. It may be challenging to select a cut-off value in the pump energy density to clearly differentiate the low-fat and high-fat males in such a case. However, the box plots in Fig. 4d based on the results in Figs. 4a and 4b indicate that there still exists biological significance among these two groups of mice since the p-value is less than 10$^{-3}$. In contrast, neither immunohistology (IHC) results of Ki-67 in Supplementary Fig. 3 nor our previous studies using stereoscope show any significance between low-fat and high-fat male colon tissues[24]. Therefore, the lasing approach may be able to detect early changes preceding colon polyps (colon cancer precursor lesions) induced by a certain dietary effect that may not be readily detected by conventional IHC or H&E methods.

The difference between the low-fat and high-fat becomes even more significant when we move to the female groups. Similar to the format used in Fig. 4, in Fig. 5a we present the lasing thresholds of 10 low-fat female mice, whose lasing thresholds have a wider range of distribution between 30 and 100 µJ/mm$^2$. In particular, we can observe that a few low-fat females (e.g., L11 and L18) have relatively low lasing thresholds, which indicate certain abnormal chromatin activities. For comparison, Fig. 5b shows the lasing thresholds of 10 high-fat female mice that are much lower and narrowly distributed within 20 to 40 µJ/mm$^2$. The histogram of all lasing thresholds collected from both low-fat and high-fat female groups is provided in Fig. 5c with the corresponding ROC curve in the inset. The area under the ROC curve is about 0.87 with a sensitivity of 0.91 at a pump energy density of 45 µJ/mm$^2$. In contrast to the male mice in Fig. 4, a cut-off pump energy density can be defined to clearly differentiate low-fat and high-fat female mice. Similar results can be validated by using the box plots in Fig. 5d, in which the calculated p-value is far less than 10$^{-3}$, showing huge biological significance between the two groups in female mice. The above result suggests that high-fat diet may cause progressive changes in chromatin condensations and higher localized DNA concentrations. Additionally, the IHC results in Supplementary Fig. 3 show that high-fat females possess higher Ki-67 expression (as quantitated in Aslam et al)[29] than low-fat females, suggesting that cell proliferation may be one of the factors for low lasing thresholds (see more discussions related to Fig. 7). It is important to note that although the samples used in Figs. 4 and 5 are regarded as normal by conventional examinations, our LEM results show otherwise. Particularly, the high-fat female group has significantly low lasing thresholds in most of the 10 mice. Therefore, the results suggest that high-fat diet may have a higher impact on the colon polyp/tumor progression and LEM can be used to detect abnormal pre-polyp nuclear changes.

In order to validate the correlation between abnormal changes in the colonic mucosa (colonic adenomas or polyps) and low lasing thresholds, we obtained the lasing thresholds from another control experiment done on 5 FFPE tissue blocks (Ad1 to Ad5) with colon adenomas. The five tissue samples were collected from 3-month old male mice treated with azoxymethane and dextran sulfate sodium (AOM/DSS) to induce adenomas (see Methods). As shown in Fig. 6a, we first investigated the lasing thresholds for the adenoma part of the tissue, which is located around the center of each tissue section (about 2 mm in radius, as shown in the inset). It is seen that the lasing threshold are relatively low in most cases, with an average between 25 $\mu J/mm^2$ to 45 $\mu J/mm^2$. In fact, the low lasing thresholds (and their narrow distributions) are very similar to our earlier findings in Figs. 5b, which attests to the correlation between the abnormal change in chromatin and the low lasing threshold. Next, in Fig. 6b we investigated the lasing thresholds for the normal part of the tissue (denoted as Ad-N1 to Ad-N5) located on the periphery of the same colonic tissue sections used in Fig. 6a. Interestingly, the lasing thresholds from those cells were even slightly lower than those from adenoma, which suggests that the abnormal chromatin deregulation may still occur in the *normal* tissue surrounding the colonic adenoma and can be detected by the lasing threshold characterization with LEM (also recall the low lasing threshold observed in Fig. 5b from the "normal" tissues of 4 mice - H11, H14, H17, and H20, which had been identified with polyp growth in other parts of their colons).

Finally, we demonstrate the rapid screening capability of LEM by imaging the colon tissues using a cut-off pump energy density of 45 $\mu J/mm^2$. In Figs. 7a and 7b we present the large-scale LEM images of low-fat and high-fat males, which are overlaid with the corresponding brightfield images. We can barely see any lasing cells in Fig. 7a, meaning that the lasing threshold of most cells are well above 45 $\mu J/mm^2$ for low-fat males. However, in Fig. 7b we see a significant number of lasing cells across the whole tissue, suggesting that high-fat induced cells have much lower lasing thresholds than low-fat. Next in Figs. 7c and 7d we present the large scale LEM images of low-fat and high-fat females, which are overlaid with the corresponding brightfield images. A few lasing cells can be seen in the low-fat female (Fig. 7c), which may indicate early development of chromatin deregulation in the cells (despite low-fat diet). In contrast, as shown in Fig. 7d, the high-fat female colon tissue has a considerably higher number of lasing cells distributed in the whole colon tissue (at both the bottom and the top of the crypts), showing an even worse chromatin deregulation scenario. This finding in females agrees well with the results in Fig. 5 that high-fat females have much lower lasing thresholds than low-fat females. To demonstrate that this is a common feature in the high-fat female colons, we provide in Fig. 7e an even larger scale LEM image of the entire colon tissue from another high-fat female. The corresponding Ki-67 IHC image is also provided for direct comparison. It is obvious that high intensity lasing cells can be detected nearly everywhere along the whole colon tissue, implying that higher chromatin deregulation may occur globally in high-fat female colon tissues. In addition, while Ki-67 has higher activities at the bottom (proliferative zones) than at the top of the crypts, the lasing cells distribute more or less equally at both the bottom and the top of the crypts across the whole colon tissue. This relatively poor correlation between Ki-67 expression and the cell lasing suggest that the lasing characteristics may reflect not only the cell proliferation, but also the number of DNA copies and chromatin condensation, etc. in a cell.

Once again, as control in Supplementary Fig. 4 we present the laser-emission imaging of the adenoma (Ad3) and its adjacent normal tissue (Ad-N3) used in Fig. 6. We can comprehend that adenoma as well as the adjacent *normal* tissues all have significant amounts of lasing cells under the same pump energy density of 45 $\mu J/mm^2$. Such highly active, abnormal nuclear changes in *normal* cells are critical indicators for early development of colon adenomas (polyps) and even tumors.

## Discussion

In this work, we used a novel imaging tool - laser emission microscopy to investigate the chromatin laser emissions of normal colon tissue from the young, 1 month old mice and compared with older 15 months old mice. As the mice grew older and were exposed to different dietary interventions, the precursor of a colon polyp may gradually transform at the biomolecular (chromatin) level in the cell nucleus. Without the ability to detect such early changes, a colon polyp may emerge and may eventually turn into an invasive cancer. Given the fact that both male and female mice were treated with either low-fat or high-fat diet, all the examined colon tissues were regarded as *normal* tissues by conventional microscope. However, we surprisingly found that colonic tissues from high-fat diet induced mice have lower lasing thresholds in both females and males as compared to corresponding low-fat mice regardless of the emergence of colon polyps at 15 months. This interesting finding implies that conventional methods may not be sufficient to detect pre-tumor lesions that are still under development at a very early stage. That is, the high-fat diet may alter, accelerate the chromatin deregulations in the cell nucleus at a very early stage, well before polyps or abnormal growth occurs, and it may have higher impact than we have previously thought. Therefore, LEM simply stands out as a potential technology to reveal the hidden DNA (chromatin) signatures during such pre-tumor development stage.

The new information may warrant that LEM would be of great significance to expand the applications in early cancer diagnosis. Here we bring up a few possibilities for the observed low lasing thresholds of laser emission. High cell proliferation may be a key factor. Although it is a well-known fact that high-fat diet will induce proliferation, with high amount of positive Ki-67 expressions, our LEM results in Fig. 7 apparently show that the location of lasing cells is not only in the base of crypts of the colon where Ki-67 are darkly stained. Second, chromatin deregulation may be another dominant factor while highly compacted chromatins will form during high DNA replication in the cell cycle. Recently, it has also become clear that the deregulation of chromatin structure plays an important role in numerous cancers[30,31].

Numerous studies including ours have demonstrated that high-fat diet alone can cause colon tumorigenesis when mice are fed with a high-fat diet over their life-span[23,24,32,33]. Therefore, this mouse model was ideal to test the lasing threshold approach to detect early changes in the colonic tissues of at-risk mice. For a better understanding of lasing mechanisms in fat-diet colon tissues, it is necessary to evaluate LEM images with IHC biomarkers of proliferation, differentiation, and apoptosis in the future. In order to translate the findings in such rodent model to human model and compare the LEM results with the changes in the possible biomarkers of colonic tissue of the subjects at risk for colon cancer, we will not only validate the LEM technology for early cancer diagnosis, but find an immediate application by providing an additional and convenient diagnostic tool to identify early changes in the colonic biopsies and to predict a future lesion in the colon that can turn into a colonic polyp if left alone. This tool can complement by supporting existing surveillance methods and assist by testing colonic tissues taken out during routine colonoscopy. Additionally, we envision that this work will open a new class of laser emission based microscopy that can be widely used in for early tumor detection in a clinical setting, as well as fundamental biological and biomedical research to detect early changes in the nuclei for better understanding at the biomolecular level regarding how colon cancer evolves.

## Methods and Materials

*Mouse model*

In this study, 50 mouse colon FFPE tissue samples were used, including 10 1-month mice (5 males, M1-M5, and 5 females, F1-F5) as baseline with no dietary intervention and 40 mice after 15 months of dietary intervention (20 males and 20 females). The design of animal experiment model is illustrated in Fig. 1a. Specifically, the 20 15-month males (labeled as L1 - L10 for low-fat and H1 - H10 for high-fat diet mice) and 20 15-month females (labeled as L11 - L20 for low-fat and H11 - H20 for high-fat diet mice) were used in a previously published paper focused on the dietary effort on colon cancer prevention[23].

The complete details regarding the mouse model and dietary interventions can be found in Ref. [[23]]. Briefly, all the 50 inbred C57BL/6 mice were obtained from Charles River, Portage, MI at 3 weeks of age. The animals were started within 1 week of arrival on either a standard low-fat rodent chow diet (AIN-76A) to serve as a control or an HFWD. HFWD was prepared according to the formulation of the Newmark stress diet and contained 20 g% fat from corn oil as compared to 5 g% in AIN-76A. The animals were euthanized at the respective time-points (1-month or 15-month) and were autopsied. Visible raised polyps were found in 4 (H11, H14, H17, and H20) out of the 20 mice at 15 months in the HFWD group. These polyps were detected grossly and verified by stereoscopic dissecting microscope. After the removal of the polyps, the colon tissues from those 50 mice were processed into FFPE blocks, part of which were subsequently stained with hematoxylin and eosin (H&E) and examined at the light microscopic level by a board-certified veterinary pathologist. The remaining unstained FFPE tissues were used in the current work (see the next section for tissue preparation for LEM).

As an additional control, FFPE samples (Ad1 to Ad5) of carcinogen-induced colon adenomatous tissue were employed. These 3 months old male mice were injected intraperitoneally with the carcinogen, azoxymethane (10 mg/kg), followed by three cycles of a 5-day course of 2% dextran sulfate sodium separated by 16 days of regular water, which induced the formation of adenomas[34]. For this experiment, mice were on a regular diet (LabDiet 5001). All of the original studies involving mice were reviewed and approved by the Institutional Animal Care & Use

Committee at the University of Michigan (Approval#: PRO00006534).

*Materials and tissue preparation*

For the current studies involving LEM, all FFPE tissue blocks were sliced into 10 μm thick sections by using a microtome (Thermo Fisher #HM355S). The selected tissue section was picked up, rinsed in 45 °C warm water-bath, and then placed on the top of a poly-L–lysine (Sigma-Aldrich #P8920) coated dielectric mirror, which was first cleaned and rinsed with lysine for better tissue adhesion. The tissues were then soaked in Xylene for 5 minutes to clean off the paraffin and wax materials. Then the tissues were soaked in different concentrations of ethanol, from 100%, 95%, 75%, to 50%, each for 2 minutes. Next, the tissues were soaked and rinsed with PBS (phosphate buffered solution, ThermoFisher # 10010023), and air dried before staining (see staining/labeling details in the next section). Finally, the tissues were covered by the top mirror (with a spacer fabricated on top of it[3]). Refractive index matching immersion oil (n = 1.42) (Thermofisher #S36937) was applied to fill the gap between the tissue and the top mirror (see details in Ref. [3]). For each tissue, approximately 10 cells within the colon tissue were randomly selected and measured.

*Staining and labeling*

For nucleic acid labeling, YO-PRO-1 Iodide solution (YOPRO, ThermoFisher #Y3603) was dissolved in PBS to form a concentration of 0.1 mM. The prepared YOPRO solution was then applied to the tissue sections for 10 minutes and rinsed with PBS solution twice before measurements.

For IHC staining, the tissue was fixed on a superfrost glass slide (ThermoScientific #15-188-48) and followed by similar process for tissue cleaning as described in the previous section. After antibody titration, immunohistochemistry was conducted using an automated system (Biocare Intellipath FLX) that enables increased throughput and reproducibility. In particular, all the tissues were incubated with 200 µL of diluted primary antibody (anti-mouse-Ki67 antibody (Abcam #16667)) for 8 hours at 4 °C. After incubation, the tissue was rinsed with PBS, followed by staining with HRP conjugated secondary antibody. Then DAB substrate solution was applied to the tissue to reveal the color of the antibody staining. After rinsing, the tissue was dehydrated through pure alcohol, then mounted with mounting medium, and finally covered with a coverslip.

*Fabry-Pérot (FP) cavity and spacer fabrication*

The FP cavity was formed by two customized dielectric mirrors (Evaporated Coating Inc., Willow Grove, PA, USA), which had a high reflectivity in the spectral range of 500-580 nm to provide optical feedback and high transmission around 473 nm for excitation. The Q-factor of the FP cavity was on the order of $10^4$ at a cavity length of 15 μm (in the absence of tissues). To ensure cavity length and system robustness, spacers were fabricated on the top mirror with a negative photoresist SU-8 on the surface of the top mirrors using standard soft lithography. The mirrors were first cleaned by solvent ultrasonication (sonicated in acetone, ethanol, and de-ionized water sequentially) and oxygen plasma treatment. Then, they were dehydrated at 150 °C for 15 minutes right before a 15 μm thick SU-8 2010 (MicroChem Corp., USA) layer was spin-coated on top. After soft-baking the SU-8-coated mirrors for 1 minute at 65 °C and 4 minutes at 95 °C, a contact lithography tool Karl Suss MA 45S was used to UV expose the mirrors through a mask with the bar-liked spacer design. The exposed mirrors were subsequently subjected to post-exposure baking at 65 °C for 1 minute and 95 °C for 5 minutes, followed by 4 minutes of development. After rinsing and drying, the SU-8 spacer on top of the mirror could be clearly seen by naked eyes. The spacers were further hard baked at 150 °C for 10 minutes and treated with oxygen plasma to improve hydrophilicity. Details of spacer can be found in Supplementary Fig. 1.

*Optical imaging system setup*

The confocal fluorescence microscopic images in Fig. 3 of colon cells were taken by using Nikon A1 Spectral Confocal Microscope with an excitation of 488 nm laser source. The bright field IHC and H&E images were taken by a high-throughput digital slide scanner (Leica Aperio® AT2). Quantitative morphometric analysis of digitized slides was performed using freely available software (Leica Aperio) using nuclear algorithm version 9.

The laser-emission images and its corresponding bright field images were captured by using a CCD (Thorlabs #DCU223C) mounted directly on top of the objective in our experimental setup (see Fig. 1b). A typical confocal setup was used to excite the sample and collect emission light from the FP cavity. In this work, a pulsed Optical parametric oscillator (OPO) laser (pulse width: 5 ns, repetition rate: 20 Hz) at 473 nm was used as the excitation source to excite the stained tissues with a laser beam size of 30 μm in diameter. The pump intensity was adjusted by a continuously variable neutral density filter, normally in the range of 10 - 200 μJ/mm$^2$. The emission light was collected through the same lens and sent to a spectrometer (Horiba iHR550, spectral resolution ~0.2 nm) for analysis.

The high-speed scanning laser-emission images were collected through the same optical setup, in which the images were taken by the CCD (20 fps, Thorlabs #DCU223C) mounted on top of the objective (NA 0.42, 20X). A high repetition rate (600 Hz) pulsed laser at 473 nm was used to pump the tissue. The raster scanning stage was home-built using two linear actuators with electric controllers (Newport #CONEX TRA25CC) and controlled by Labview. The high-speed scanning was conducted by a continuous linear scanning method with a step size of 10 μm. For a 1 mm x 1 mm area, only 200 seconds are required to form an LEM image. Currently, the largest mapping area is approximately 3 mm x 3 mm.

**Data availability**

All raw images and data generated in this work, including the representative images provided in the manuscript, are available from the corresponding author upon request. The authors declare that all data supporting the findings of this study are available within the paper and its supplementary information.

**Acknowledgments**

We acknowledge the support from the National Science Foundation (ECCS-1607250) and ULAM at the University of Michigan for the IHC service.

**Author contributions**

Y.C., M.N.A., and X.F. conceived the research and designed the experiments; Y.C. and Q.C. performed the experiments; Q.C. fabricated the spacer; Y.C. and G.C. prepared the samples; Y.C., X.T., M.N.A., I.B., and X.F. analyzed data; Y.C., M.N.A., and X.F. wrote the paper.

**Additional information**

The authors declare no competing financial interests.

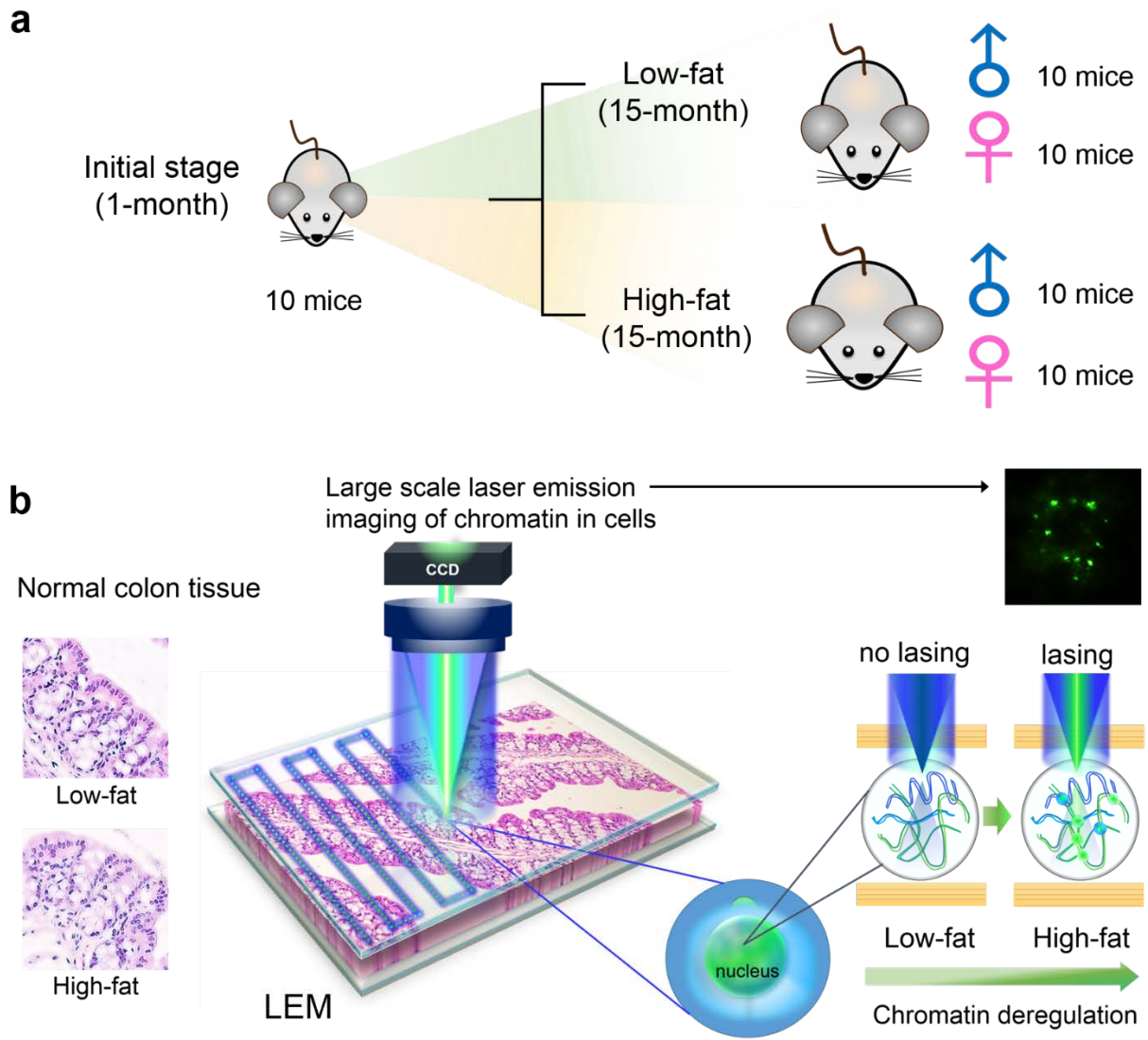

**Figure 1. Conceptual illustrations of the experimental design and setup. a,** Experimental animal model used in this work. We focused on the precursor of very early stage development of adenomas by using mouse colon tissues as a model system. Formalin fixed paraffin-embedded (FFPE) tissues from 50 mice were examined, including 10 1-month mice (5 males and 5 females), 40 15-month mice (20 males and 20 females with low-fat and high-fat dietary treatment). **b,**

Conceptual illustration of a large-scale rapid scanning laser-emission microscopy (LEM) for studying cancer precursors, in which an FFPE tissue is sandwiched within a high-Q Fabry-Pérot (FP) cavity. Tissue thickness for all samples used in LEM was 10 μm. Excitation wavelength = 473 nm. Details can be found in Supplementary Fig. 1. The left panel shows that no significant morphological difference is observed in the H&E images between a low-fat colon tissue and a "apparently normal" high-fat colon tissue. The right panel shows that significant differences can be seen in the LEM images (and other lasing characteristics) between the same two tissues in the left panel. For example, the high-fat tissue has a lower lasing threshold, and more lasing cells and stronger lasing emission under the same external excitation than the low-fat tissue. Those differences reflect the abnormal cell proliferation and chromatin deregulation (condensation) in the high-fat colon tissue, despite the absence of abnormal morphological changes in the H&E image.

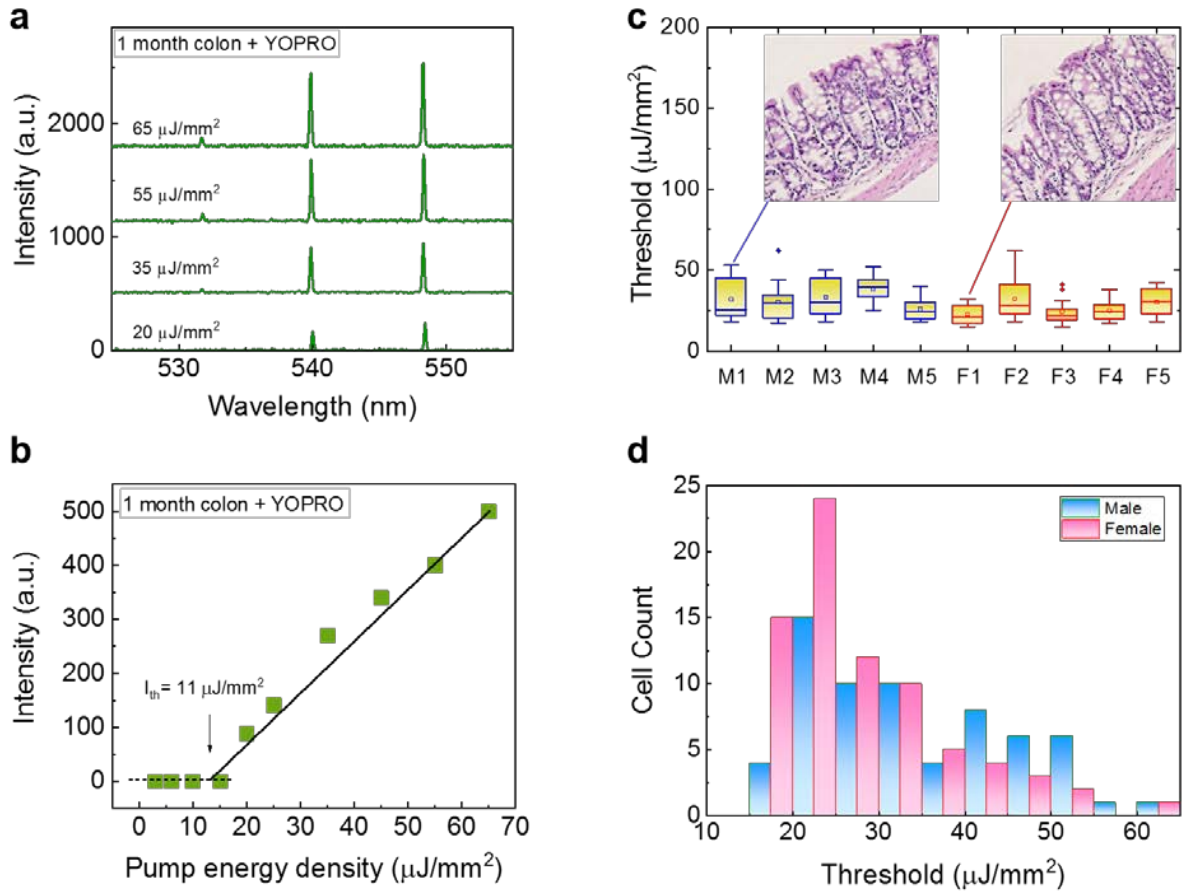

**Figure 2. Lasing properties of 1-month mice. a,** Examples of lasing spectra of a normal colon tissue from a 1-month old mouse (initial stage) stained with YOPRO (0.1 mM) under various pump energy densities. All curves are vertically shifted for clarity. **b,** Spectrally integrated laser output as a function of pump energy density extracted from the laser spectra of the normal colon tissue in **a**. The solid line is the linear fit above the lasing threshold, which was 11 μJ/mm². Tissue thickness = 10 μm. Cavity length = 15 μm. Excitation wavelength = 473 nm. **c**, Statistics of the lasing threshold for cells in the normal colon tissues stained with YOPRO from five 1-month old male mice (labeled as M1-M5) and five 1-month old female mice (labeled as F1-F5). Exemplary H&E images of male and female mice colon tissue are provided in the insets. **d**, Histogram of all male/female normal colon cell lasing thresholds (N = 100) extracted from **c**.

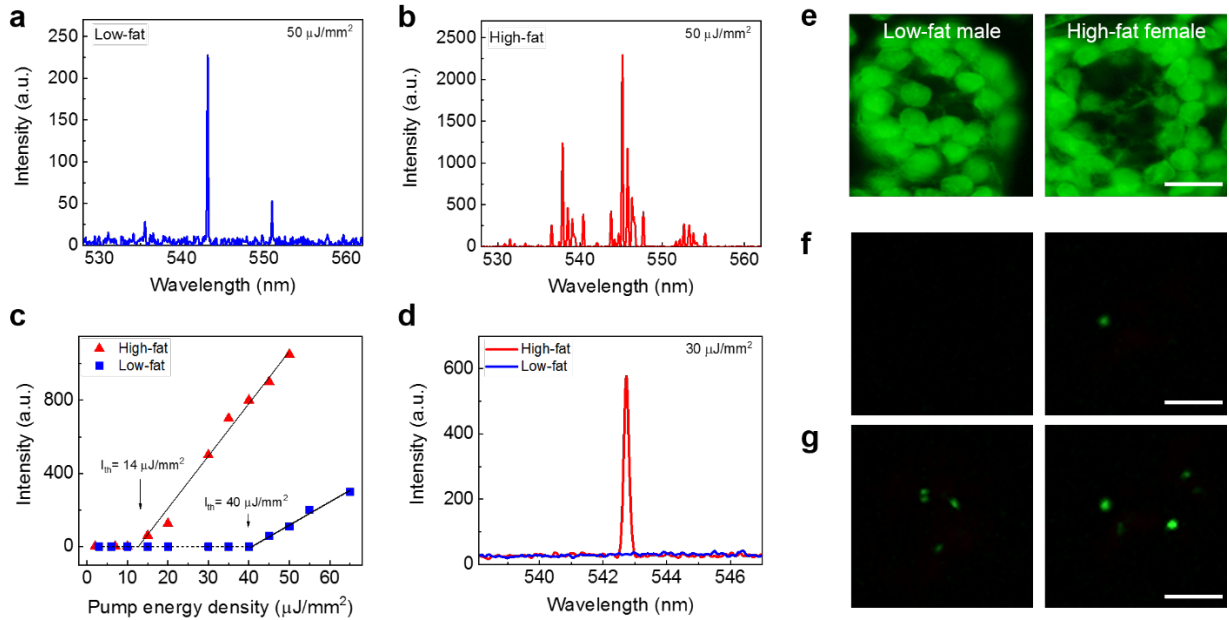

**Figure 3. Comparison of laser emissions between low-fat and high-fat induced colon tissue.
a,** Example of lasing spectrum of a normal colon tissue from a 15 months old low-fat male mouse stained with YOPRO (0.1 mM) under a pump energy density of 50 µJ/mm$^2$. **b,** Example of lasing spectrum of a normal colon tissue from a 15 months old female high-fat mouse stained with YOPRO (0.1 mM) under a pump energy density of 50 µJ/mm$^2$. **c,** Spectrally integrated laser output as a function of pump energy density extracted from the laser spectra of the low-fat (blue squares) and high-fat (red triangles) colon tissues in **a** and **b**. The solid lines are the linear fit above the lasing threshold, which was 14 µJ/mm$^2$ and 40 µJ/mm$^2$ for the high-fat mouse and low-fat mouse, respectively. Tissue thickness = 10 µm. Cavity length = 15 µm. Excitation wavelength = 473 nm. **d,** Direct comparison of laser emission spectra of the same low-fat male (blue curve) and high-fat female (red curve) colon tissues used in **c** at fixed pump energy density of 30 µJ/mm$^2$. **e,** Representative fluorescence microscopic images of colon cells from the same low-fat male (left column) and high-fat female (right column) colon tissues used in **d**. **f,** Representative laser emission images of colon cells from the same low-fat male (left column) and high-fat female (right column) mice used in **e** when pumped at 30 µJ/mm$^2$. **g,** Laser emission images of the same sets of cells in **f** when the pump energy density increased to 70 µJ/mm$^2$. All scale bars, 20 µm.

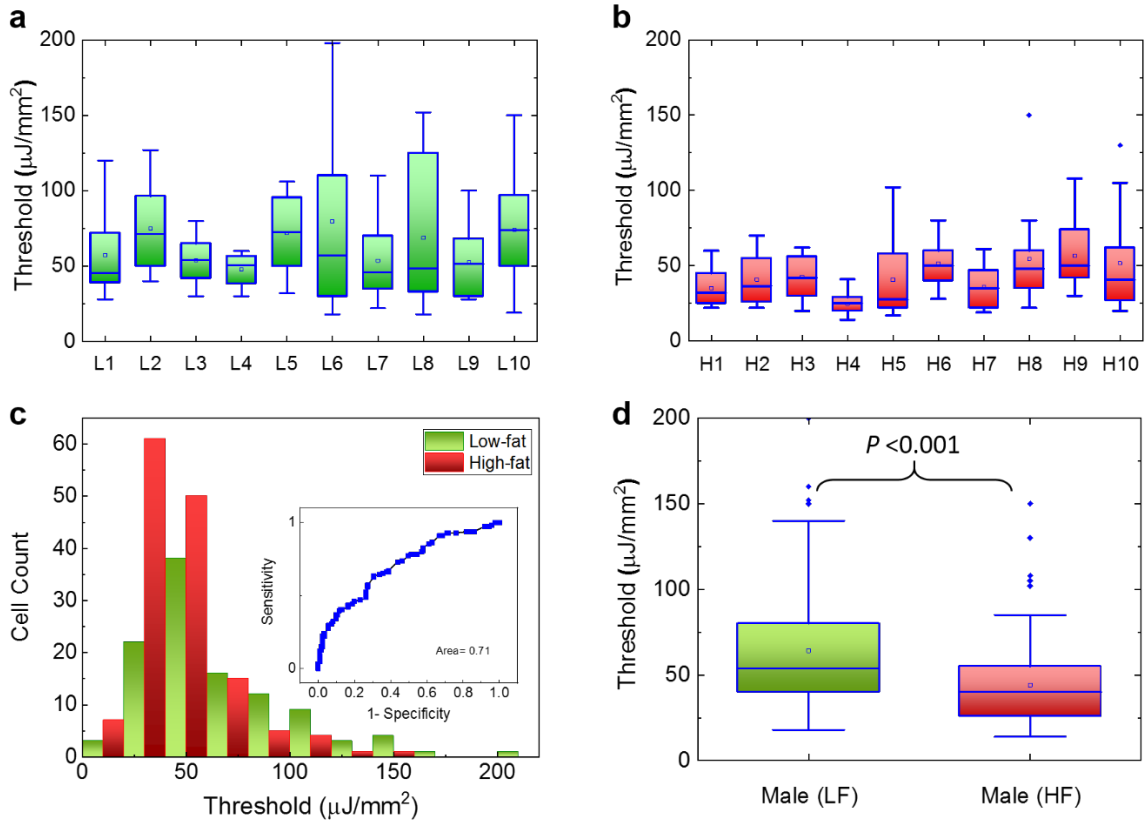

**Figure 4. Assessment of the lasing threshold of low-fat and high-fat induced male mice. a**, Statistics of the lasing threshold for cells in individual low-fat male colon tissues (10 mice) stained with YOPRO. **b**, Statistics of the lasing threshold for cells in individual high-fat male colon tissues (10 mice) stained with YOPRO. **c**, Histogram of all low-fat/high-fat cell lasing thresholds (N = 230) extracted from **a** and **b**. The inset is the corresponding Receiver Operating Characteristics (ROC) curve. The ROC curve is plotted by using the different excitations (in units of µJ/mm$^2$) as the cut-off criterion. The area under the fitted curve is 0.71 and the sensitivity of 50% is obtained based on the cut-off criterion of 40 µJ/mm$^2$. **d**, Statistic comparison of the lasing threshold for all 230 cells extracted from **a** and **b**. The p-value between the two groups is <10$^{-3}$. All tissues were 10 µm in thickness and were sandwiched in a 15 µm long cavity. [YOPRO] = 0.1 mM for all 20 mice samples.

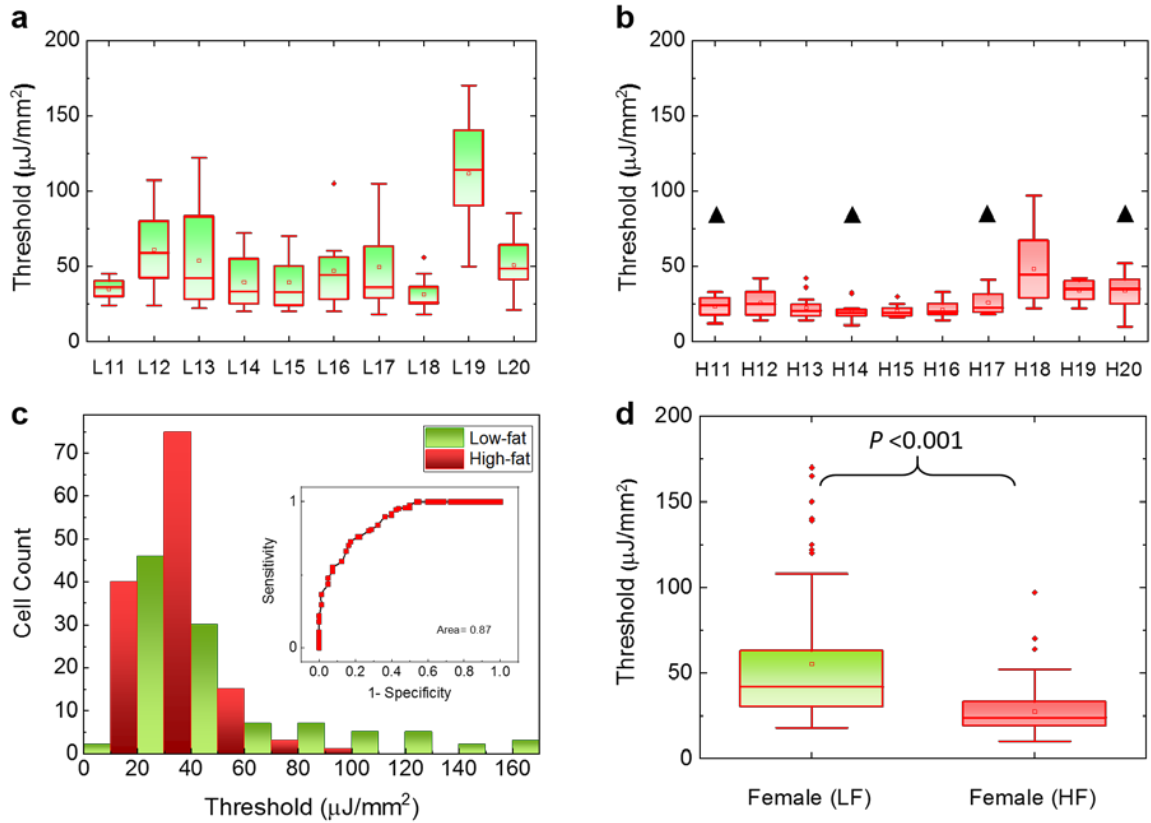

**Figure 5. Assessment of the lasing threshold of low-fat and high-fat induced female mice. a**, Statistics of the lasing threshold for cells in individual low-fat female colon tissues (10 mice) stained with YOPRO. **b**, Statistics of the lasing threshold for cells in individual high-fat female colon tissues (10 mice) stained with YOPRO. The black triangles above the 4 data sets indicate the 4 mice that had been identified with polyp growth in other parts of their colons (not in the parts used in the current study). **c**, Histogram of all low-fat/high-fat cell lasing thresholds (N = 204) extracted from **a** and **b**. The inset is the corresponding Receiver Operating Characteristics (ROC) curve. The ROC curve is plotted by using the different excitations (in units of µJ/mm$^2$) as the cut-off criterion. The area under the fitted curve is 0.87 and the sensitivity of 90% is obtained based on the cut-off criterion of 40 µJ/mm$^2$. **d**, Statistic comparison of the lasing threshold for all 204 cells extracted from **a** and **b**. The p-value between the two groups is $\ll 10^{-3}$. All tissues were 10 µm in thickness and were sandwiched in a 15 µm long cavity. [YOPRO] = 0.1 mM for all 20 mice samples.

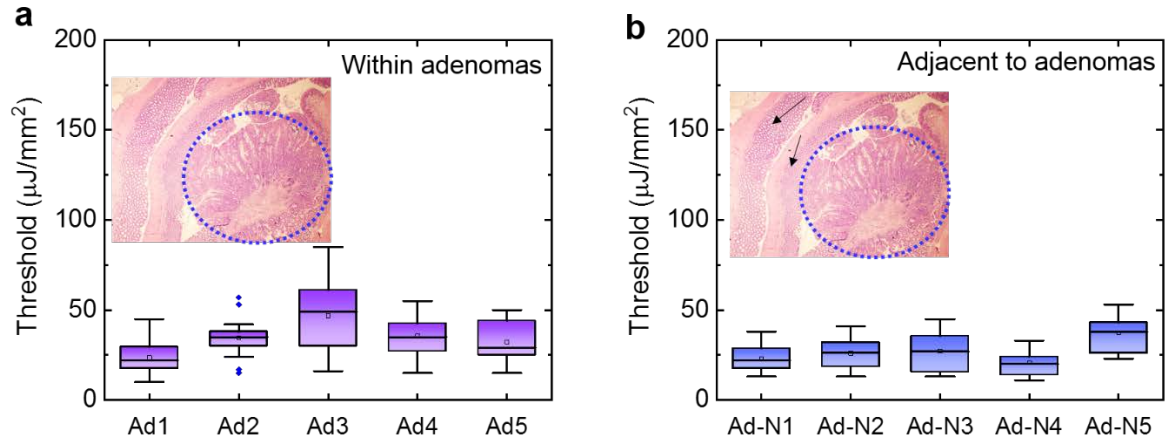

**Figure 6. Assessment of the lasing threshold of adenomas and adjacent tissues. a**, Statistics of the lasing threshold for cells in individual colon adenoma tissues (5 3-month male mice on regular diet, denoted as Ad1 to Ad5) stained with YOPRO. The inset shows an exemplary H&E image of colon tissue. The blue dotted circle (2 mm in radius) indicates the area that has colon adenoma. **b**, Statistics of the lasing threshold for cells in the normal colon tissues adjacent to adenoma from the exactly same tissue section in **a** stained with YOPRO (Ad-N1 to Ad-N5 denote the adjacent normal tissues from Ad1 to Ad5 used in **a**, respectively). The black arrows in the inset H&E image show the adjacent normal tissue used for measurements.

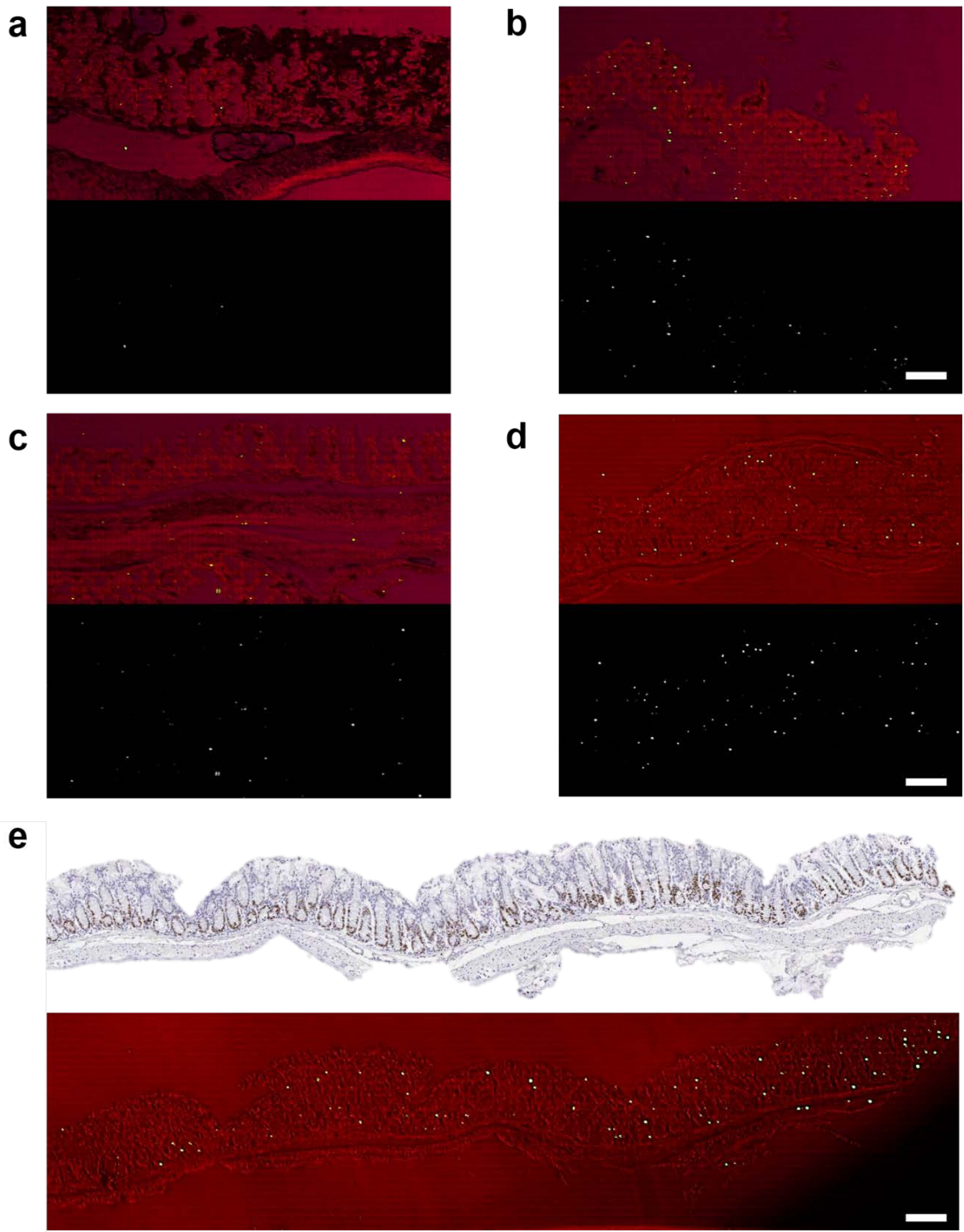

**Figure 7. Laser-emission imaging of mice colon tissues. a,** Representative low-fat male laser emission image (bottom) that is overlaid with the corresponding brightfield image (top). **b,** Representative high-fat male laser emission image (bottom) that is overlaid with the corresponding

brightfield image (top). **c,** Representative low-fat female laser emission image (bottom) that is overlaid with the corresponding brightfield image (top). **d,** Representative high-fat laser emission image (bottom) that is overlaid with the corresponding brightfield image (top). **e,** IHC microscopic image of a high-fat female colon tissue labeled with Ki-67 biomarker (top image). The corresponding laser emission image (large scale scanning) overlaid with the corresponding brightfield image is shown in the bottom image. The tissues for the top image and the bottom image were sliced from the same FFPE block. The total area is 3 mm x 0.5 mm. All images were scanned and measured under a pump energy density of 45 µJ/mm$^2$. All scale bars, 100 µm. The estimated number of lasing cells for images in **a, b, c, d,** and **e** are 4, 45, 16, 64, and 125, respectively.

# Supplementary Information for

# Chromatin Laser Imaging Reveals Abnormal Nuclear Changes for Early Cancer Detection

Yu-Cheng Chen[1,2], Qiushu Chen[1], Xiaotain Tan[1], Grace Chen[3], Ingrid Bergin[4], Muhammad Nadeem Aslam[5*], and Xudong Fan[1*]

[1]Department of Biomedical Engineering, University of Michigan,
1101 Beal Ave., Ann Arbor, MI, 48109, USA

[2]School of Electrical and Electronics Engineering, Nanyang Technological University,
50 Nanyang Ave, 639798, Singapore

[3]Comprehensive Cancer Center, University of Michigan, Ann Arbor, MI 48109, USA

[4]Unit for Laboratory Animal Medicine, University of Michigan, Ann Arbor, MI 48109, USA

[5]Department of Pathology, University of Michigan,
1301 Catherine Street, Ann Arbor, MI 48109, USA

*Correspondence: xsfan@umich.edu, mnaslam@med.umich.edu

# 1. Design of the FP cavity with SU-8 spacer

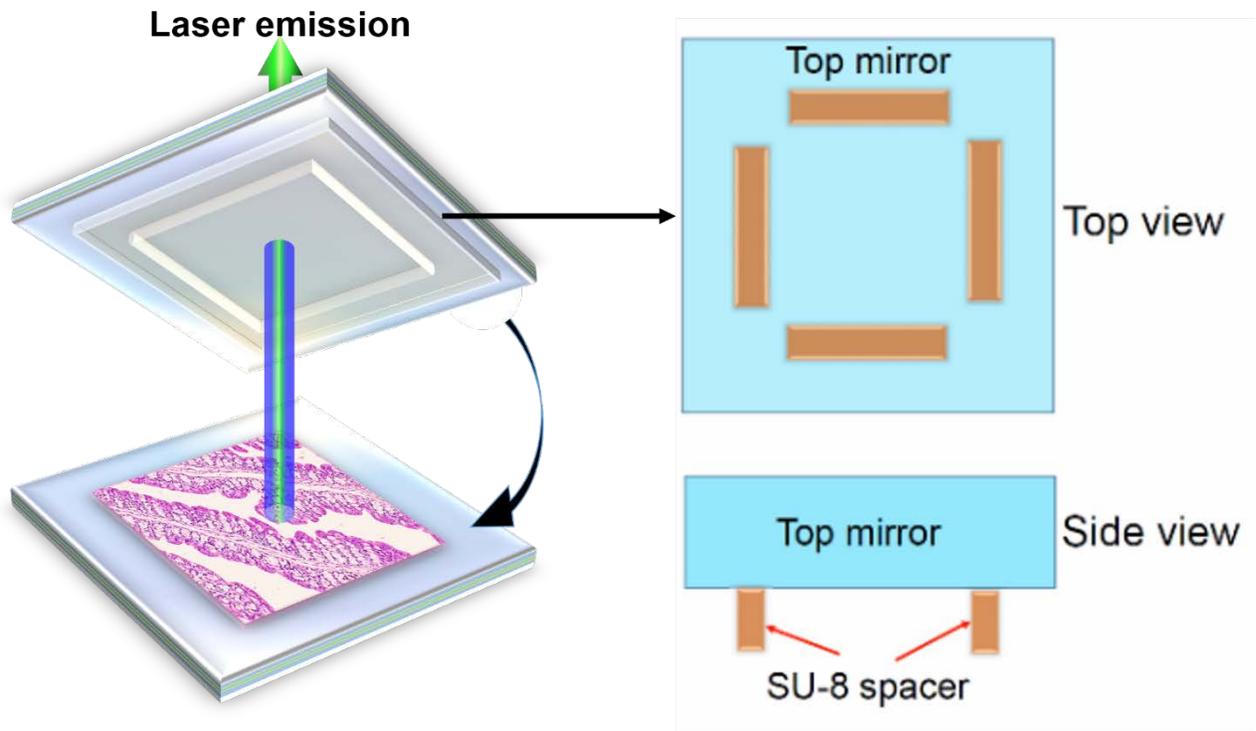

**Figure S1.** (Left panel) FP cavity is formed by a bottom mirror on which a colon tissue section (10 µm in thickness) is placed and a top mirror on which SU-8 spacers (height: 15 µm) are microfabricated. The spacer height determines the FP cavity length. (Right panel) Illustrations of the top mirror with spacers.

**2. Exemplary H&E images used in Figs. 4 and 5**

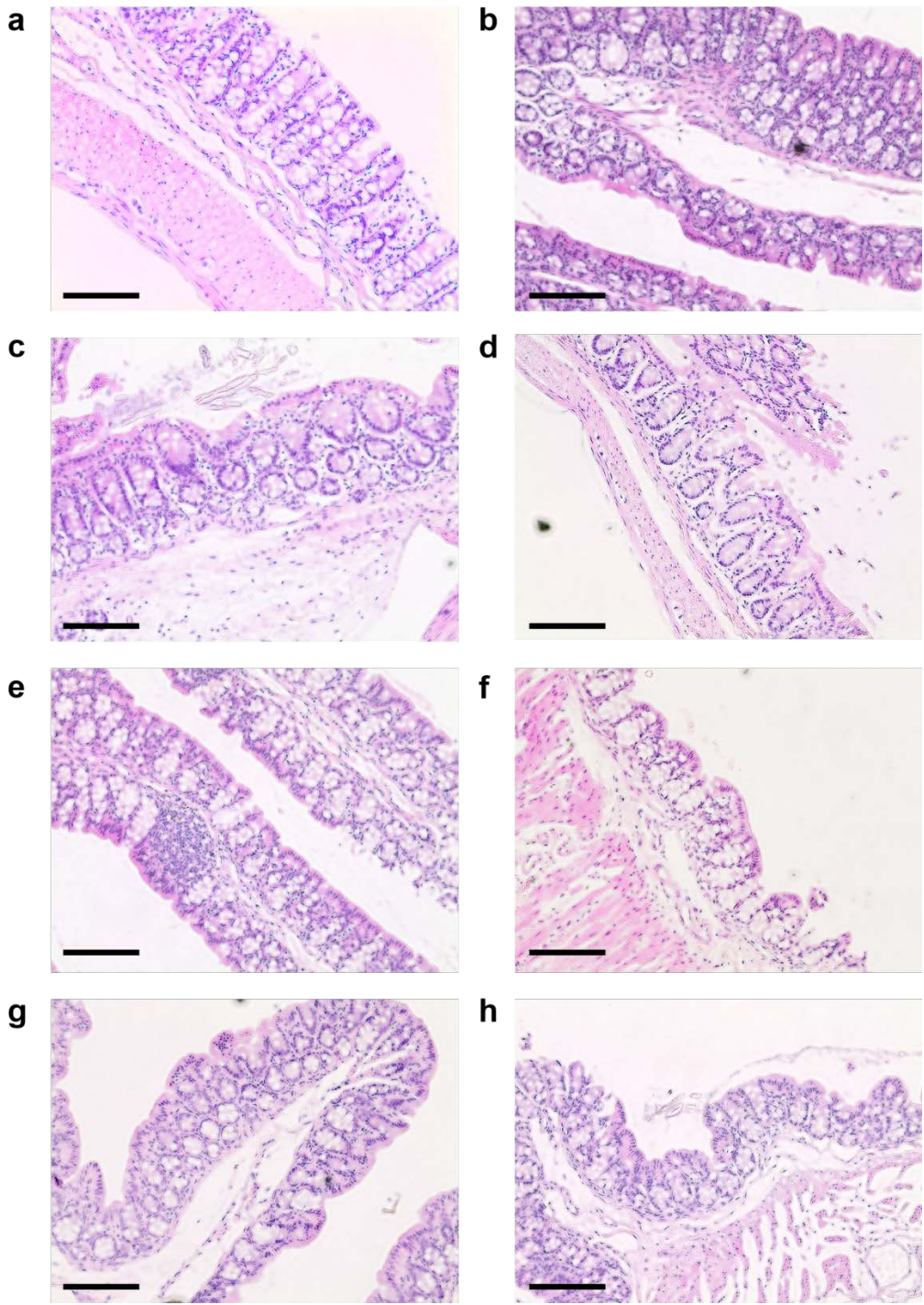

**Figure S2. a-b,** H&E images of a low-fat male colon tissue (a) L1 and (b) L2. **c-d,** H&E images of a high-fat male colon tissue (c) H3 and (d) H10. **e-f,** H&E images of a low-fat female colon tissue (e) L16 and (f) L17. **g-h,** H&E images of a high-fat female colon tissue (g) H15 and (h) H16. All tissues were examined and diagnosed by pathologists. All scale bars, 100 µm.

## 3. IHC results of high-fat and low-fat colon tissues labeled with Ki-67

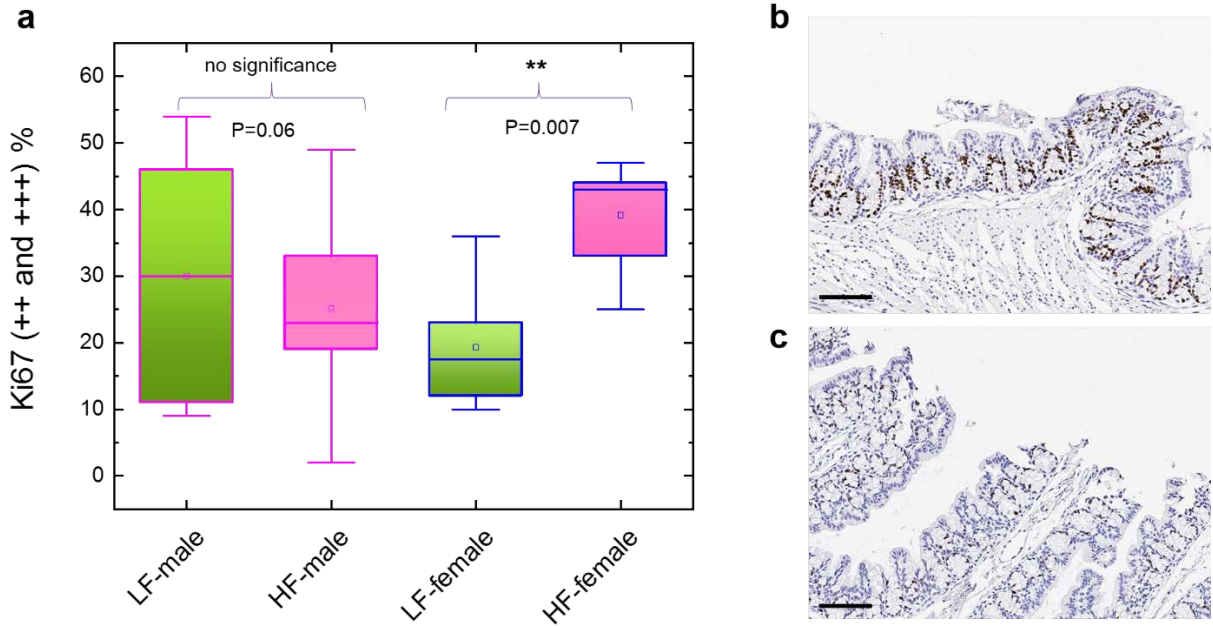

**Figure S3. a,** Comparison of Ki-67 proliferation biomarker expression among the low-fat male, high-fat male, low-fat female, and high-fat female groups. In each group 5 mice were randomly selected and used nuclear algorithm to analyze the Ki-67 expressions. The Ki-67 expressions were calculated based on the percentage of 2+ and 3+ positive Ki-67 cells in the colon tissue section. Details of the analyzing protocol can be found in the Experimental Section in the main text and reference [29]. **b,** Exemplary IHC images of a high-fat female mouse with Ki-67 expression shown in dark brown color. **c,** Exemplary IHC images of a low-fat female mouse with Ki-67 expression shown in dark brown color. Scale bars, 100 µm.

## 4. Laser emission images of adenoma and adjacent tissues

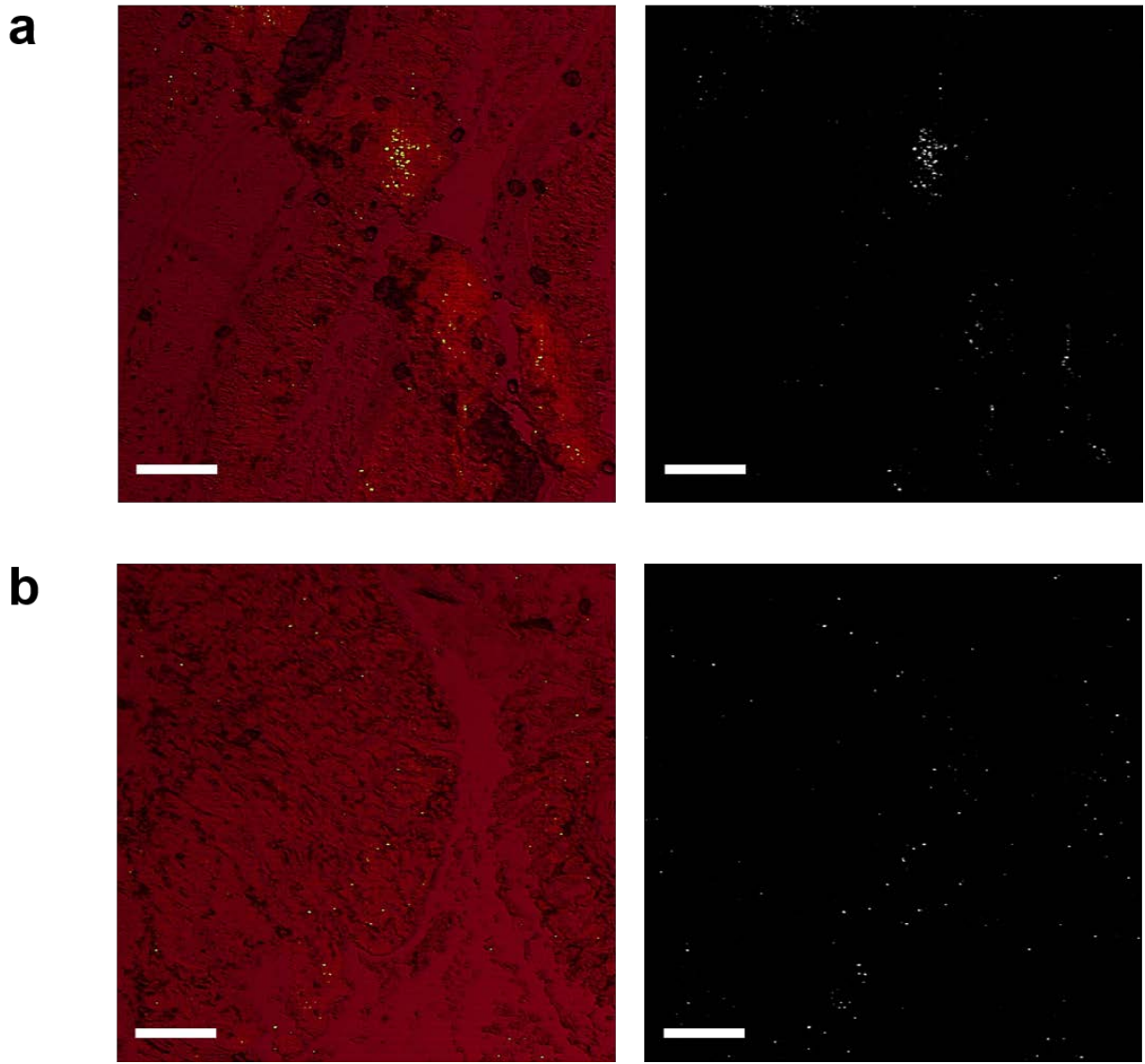

**Fig. S4. a,** (Left) Representative laser emission image of an adenoma tissue overlaid with the corresponding brightfield image. (Right) The laser emission image alone. **b,** (Left) Representative laser emission image of a normal tissue adjacent to adenoma overlaid with the corresponding brightfield image. (Right) The laser emission image alone. The LEM images were scanned and measured under a pump energy density of 45 µJ/mm$^2$. Scale bars, 200 µm.